\documentclass[%
 reprint,
 amsmath,amssymb,
 aps,
]{revtex4-2}

\usepackage{graphicx}
\usepackage{dcolumn}
\usepackage{bm}
\usepackage{float}
\usepackage[many]{tcolorbox}
\usepackage{lipsum}
\usepackage{amsmath,cancel}

\usepackage{shuffle}
\usepackage{graphbox}
\usepackage{environ}
\usepackage{physics}
\usepackage{tabularx}
\usepackage{MnSymbol}
\begin{document}

\title{All-loop geometry for four-point correlation functions}

\author{Song He$^{1,2,3}$\footnote{songhe@itp.ac.cn}, Yu-tin Huang$^{4,5}$\footnote{yutinyt@gmail.com}, Chia-Kai Kuo$^{4,6}$\footnote{chiakaikuo@gmail.com}}
\affiliation{
$^{1}$CAS Key Laboratory of Theoretical Physics, Institute of Theoretical Physics, Chinese Academy of Sciences, Beijing 100190, China \\
$^{2}$School of Fundamental Physics and Mathematical Sciences, HIAS-UCAS;\\
International Centre for Theoretical Physics Asia-Pacific, Beijing/Hangzhou, China\\
$^{3}$Peng Huanwu Center for Fundamental Theory, Hefei, Anhui 230026, P. R. China\\
$^{4}$Department of Physics and Center for Theoretical Physics, National Taiwan University, Taipei 10617, Taiwan\\
$^{5}$Physics Division, National Center for Theoretical Sciences, Taipei 10617, Taiwan\\
$^{6}$Max–Planck–Institut für Physik, Werner–Heisenberg–Institut, D–85748 Garching bei München, Germany
}\date{\today}

\begin{abstract}
In this letter, we consider a positive geometry conjectured to encode the loop integrand of four-point stress-energy correlators in planar $\mathcal{N}=4$ super Yang-Mills. Beginning with four lines in twistor space, we characterize a positive subspace to which an $\ell$-loop geometry is attached. The loop geometry then consists of $\ell$ lines in twistor space satisfying positivity conditions among themselves and with respect to the base. Consequently, the $\textit{loop geometry}$ can be viewed as fibration over a $\textit{tree geometry}$. The fibration naturally dissects the base into chambers, in which the degree-$4 \ell$ loop form is unique and distinct for each chamber. Interestingly, up to three loops, the chambers are simply organized by the six ordering of $x^2_{1,2}x^2_{3,4}$, $x^2_{1,4}x^2_{2,3}$ and $x^2_{1,3}x^2_{2,4}$. We explicitly verify our conjecture by computing the loop-forms in terms of a basis of planar conformal integrals up to $\ell=3$, which indeed yield correct loop integrands for the four-point correlator.

\end{abstract}
\maketitle


\section{Introduction}

The Amplituhedron~\cite{Arkani-Hamed:2013jha, Arkani-Hamed:2013kca, Arkani-Hamed:2017vfh} is a collection of positive geometries whose canonical forms~\cite{Arkani-Hamed:2017mur} are conjectured to give all-loop integrands for scattering amplitudes in planar ${\cal N}=4$ supersymmetric Yang-Mills theory (SYM). Surprisingly, a simple deformation of its definition gives the so-called ABJM Amplituhedron~\cite{He:2023rou, He:2022cup}, which achieves the same for ABJM theory~\cite{Aharony:2008ug, Hosomichi:2008jb}. A nice by-product in~\cite{He:2023rou} is a new way for computing canonical forms of these geometries: loop geometry can be viewed as fibration over the tree geometry which naturally dissects the latter into chambers. It is then very natural to ask whether we can extend the amplituhedron, as well as this new way of computing loop-forms based on chambers, to other observables. A promising prospect would be the correlation functions of stress-energy multiplets~\cite{Eden:2011we} in ${\cal N}=4$ SYM, which have been significant objects of great interests (c.f. a recent review~\cite{Heslop:2022xgp} and references therein).  In~\cite{Eden:2017fow}, a geometric object called the {\it correlahedron} has been introduced as certain ``off-shell" generalization of the Amplituhedron, which is conjectured to encode correlation functions: via the remarkable duality the correlator gives the square of amplitudes/Wilson loops in lightlike limits~\cite{Eden:2010ce, Eden:2011ku, Alday:2010zy, Eden:2010zz, Eden:2011yp, Adamo:2011dq}, where the correlahedron is supposed to reduce to (the square of) the amplituhedron~\cite{Dian:2021idl}!

In this letter, we focus on four-point correlator: we consider a positive geometry and provide strong evidence that its canonical form gives the correct loop integrand. The integrand for the four-point correlator has been determined to ten loops~\cite{Bourjaily:2016evz, Bourjaily:2015bpz}: due to the discovery of a hidden permutation symmetry, the basis for integrands is reduced to enumerating planar graphs with certain conformal properties, or the  so-called $f$ graphs~\cite{Eden:2011we}; an efficient graphical method has been developed for obtaining the coefficients of such graphs which gives compact formulas for the correlator integrand. 

Here we will focus on the super correlator naturally described as a potential in chiral superspace~\cite{Chicherin:2014uca}. Similar to the correlahedron, we define our geometry in bosonized twistor space, where a set of positivity conditions are imposed on four lines representing the positions of the operators. In position space, these positivity conditions simply translate to $x^2_{i,j}>0$, where $x_i$s are the position of the operators. This subspace is two-dimensional and we introduce a coordinate chart that is a double cover. On each point of this ``tree kinematics", we will attach additional $\ell$ pairs of twistors at $\ell$ loops. Requiring each pair to be mutually positive and with respect to the point it is attached to, the resulting $4\ell$-dimensional subspace defines our loop geometry. Thus the full geometry consists of a loop geometry fibered over the tree geometry and the total dimension is $2+4\ell$-dimensional.  

We compute the canonical forms on this geometry, which is defined to have logarithmic singularities at its boundaries. Since each point in kinematic space has its own loop-geometry, the tree region can be naturally dissected into \textit{chambers}, for which the loop-form is the same within one chamber and differs once one crosses the boundary. Up to three loops, we find that there are 6 chambers associated with $6$ different ordering of $x^2_{1,2}x^2_{3,4}$, $x^2_{1,4}x^2_{2,3}$ and $x^2_{1,3}x^2_{2,4}$.

\section{The definition}
Let us begin with the definition of the geometry. We collect $4$ bitwistors $X^{IJ}_i=Z^I_{[1,i}Z^{J}_{2],i}$ where $I,J=1,\cdots,4$ and $i=1,\cdots,4$, representing the 4 positions into a $4\times 8$ matrix $\{X_1, X_2, X_3,X_4\}$. We consider the subregion defined through the following non-negative condition 
\begin{equation}\label{eq: TreeRegion0}
 \langle i,j\rangle\equiv \frac{1}{4!}\epsilon_{IJKL}X_i^{IJ}X_j^{KL}>0\,.
\end{equation}
The bitwistors and position space are related via
\begin{equation}
x^2_{i,j}\equiv(x_i-x_j)^2=\frac{\langle i,j\rangle}{\langle i, I\rangle\langle j, I\rangle}
\end{equation}
where $I$ is the infinity twistor that breaks the SL(4) conformal symmetry. Note that the constraint in eq.\eqref{eq: TreeRegion0} is invariant under a $\mathfrak{g}=$ left GL(4)$\times$ right (GL(2))$^4$ action on the bitwistors. We can utilize this freedom to choose the frame where $\{X_1, X_2, X_3,X_4\}$ takes the form,  
\begin{equation}\label{eq: GaugeFix}
\begin{pmatrix}
\mathbb{I}_{2\times 2} & 0 & \mathbb{I}_{2\times 2} & \mathbb{I}_{2\times 2}\\
0 & \mathbb{I}_{2\times 2} & \mathbb{I}_{2\times 2} & \begin{matrix}
c_1 & 0 \\
0 & c_2
\end{matrix}
\end{pmatrix}\,,
\end{equation} 
with $c_1, c_2$ related to the cross-ratios as  $(1-c_1)(1-c_2)=\frac{\langle 1,2\rangle \langle 3,4\rangle}{\langle 1,3\rangle \langle 2,4\rangle}=v$  and $c_1 c_2=\frac{\langle 1,4\rangle \langle 2,3\rangle }{\langle 1,3\rangle \langle 2,4\rangle}=w$ (they are conventionally denoted as $z, \bar{z}$). Note that this is a double cover of eq.(\ref{eq: TreeRegion0}) with $c_1\leftrightarrow c_2$ yielding the same point.

In this two-dimensional subspace, we will further impose the constraint $\Delta^2=s^2{+}t^2{+}u^2{-}2(st{+}tu{+}us)>0$, with $s\equiv \langle 1,2\rangle\langle 3,4\rangle$, $t\equiv \langle 1,4\rangle\langle 2,3\rangle$ and $u\equiv \langle 1,3\rangle\langle 2,4\rangle$. To motivate this constraint, note that in the frame eq.(\ref{eq: GaugeFix}), $\Delta^2=(c_1{-}c_2)^2$. Thus the constraint is equivalent to requiring $c_{1,2}$ being real. We will define the tree-region as the union between eq.(\ref{eq: TreeRegion0}) and $\Delta^2>0$,
\begin{equation}\label{eq: TreeRegion}
\mathbb{T}_4: \quad \langle i,j\rangle>0 \cup \Delta^2>0\,.
\end{equation}
Now at each point in $\mathbb{T}_4$ we can introduce a loop-region by introducing $\ell$ pair of twistors $(Z_{A_{\ell_i}}, Z_{B_{\ell_i}})$ with $i=1,\cdots, \ell$ at $\ell$ loops. Again modding GL(2) this introduces an additional four-degrees of freedom for each loop, and the subregion is defined via  
\begin{equation}\label{eq: LoopRegion}
\langle (A_{\ell_i},B_{\ell_i}),X_j\rangle>0, \quad \langle (A_{\ell_i},B_{\ell_i}),(A_{\ell_j},B_{\ell_j})\rangle>0\,.
\end{equation}
We define the loop-geometry as the canonical form of this subregion. Thus the total geometry is $2{+}4\ell$-dimensional. Again since the loop-form may depend its ``base point" in $\mathbb{T}_4$, this suggests that $\mathbb{T}_4$ should be dissected into \textit{chambers}~\cite{He:2023rou}, with each chamber defined as regions in $\mathbb{T}_4$ where the loop form is uniform. Note that for $\Delta^2<0$, the loop form actually vanishes identically which is why $\mathbb{T}_4$ is constrained to have $\Delta^2>0$.

We will be constructing the canonical form of $\mathbb{T}_4$ or of its chambers. Using the parametrization in eq.(\ref{eq: GaugeFix}) one constructs a two-form $f(c_1, c_2) dc_1dc_2$ that has logarithmic singularity at the boundary of eq.(\ref{eq: TreeRegion0}) and that of the chambers. The canonical form is then uplifted into a $\mathfrak{g}$ invariant form by introducing 
$Y_\alpha^{\mathcal{I}}\in$ Gr($4,8$) where 
\begin{equation}
Y_\alpha^{\mathcal{I}}=\sum_{i=1}^8C_{\alpha,i }\cdot \mathbf{X}_i^{\mathcal{I}}\,.
\end{equation}
with $\mathbf{X}=\{Z_{1,1},Z_{1,2},\cdots,Z_{4,1},Z_{4,2}\}$. 
The original $X$ can then be identified as $X_i=Y^\perp\cdot\mathbf{X}_i
$ and
\begin{equation}\label{eq: TwistorDef}
\langle i, j\rangle =\llangle Y, \mathbf{X}_i, \mathbf{X}_j\rrangle 
\end{equation}
where $\llangle \cdots \rrangle$ takes the determinant of $8\times 8$ matrix. Then the canonical form is uplifted by identifying $dc_1 dc_2 =\frac{d\mu_Y}{\Delta^2}$ with $d\mu_Y\equiv\prod_{\alpha=1}^4\llangle Y d^4Y_\alpha\rrangle$ and:
\begin{equation}
f^\pm_\sigma(c_1,c_2)\,dc_1dc_2=\omega^\pm_\sigma(Y,\mathbf{X})\,d\mu_Y\,.
\end{equation}
where $\sigma$ labels the distinct chambers and $\,^\pm$ indicates the two branches of the double cover ($c_1>c_2$ and $c_2>c_1$).

For the loop-form, one simply expand the loop twistors $(A_{\ell_i}, B_{\ell_i})$ on any four external twistors $Z$s and compute the $4\ell$-form that is the canonical form for eq.\eqref{eq: LoopRegion}, which we denote as $\Omega^{(\ell) \pm}_\sigma$. The potential for the correlation function is then identified as $\mathcal{G}(\mathbf{X},Y)$ through
\begin{equation}
\mathcal{G}(\mathbf{X},Y)=\sum_{\sigma, \pm} \;\omega^\pm_\sigma (Y,\mathbf{X})\,\Omega^{(L) \pm}_\sigma \,.
\end{equation}
Finally as $\mathcal{G}(\mathbf{X},Y)$ is itself conformal invariant, the infinity twistor in eq.\eqref{eq: TwistorDef} will cancel and one simply replace  $\langle i, j \rangle\rightarrow x^2_{i,j}$ to recover the correlator in position space. 

Before moving on to explicit forms, we comment that the definition of the geometry in eq.(\ref{eq: TreeRegion}) combined with eq.(\ref{eq: LoopRegion}) enjoys an $P_{4{+}\ell}$ permutation invariance. While our triangulation in terms of different chambers obscures this property, it will emerge in the sum.

\section{The tree geometry}
We first consider the form associated with the tree geometry $\mathbb{T}_4$. Since the region is determined by six inequalities $\langle i, j\rangle>0$, it is straightforward to obtain the canonical form using the parametrization in eq.(\ref{eq: GaugeFix}). The form associated with $c_1>c_2$ is given as, 
\begin{equation}
\frac{(1 {-}c_2 {+} c_1 c_2)dc_1 dc_2}{(c_1{-}1)(c_2{-}1)c_1 c_2},
\end{equation}
and the form for $c_2>c_1$ is obtained by simply exchanging $1\leftrightarrow 2$. The two forms are then uplifted to  
 \begin{equation} \label{eq:tree-form}
\omega^\pm(Y,\mathbf{X})=\left( \frac{s{+}t{+}u\pm \Delta}{stu}\right)\frac{\langle \! \langle \mathbf{X}_1, \mathbf{X}_2, \mathbf{X}_3, \mathbf{X}_4\rangle \! \rangle^4}{\Delta^2} \,.
\end{equation} 
Note that this does not give the tree-level correlator. This is not a surprise given that the loop-level correlators are not proportional to the tree correlator. The prefactor $\langle\! \langle \mathbf{X}_1, \mathbf{X}_2, \mathbf{X}_3, \mathbf{X}_4\rangle\!\rangle^4$ is necessary to balance the conformal weight at each point, and will be universal for all-loop geometries. This turns out to give the universal prefactor $R(1,2,3,4)$ for the four-point correlator as we show in the supplementary material~A.

\section{One and two-loops}
\begin{figure}[H]
\includegraphics[width=8cm]{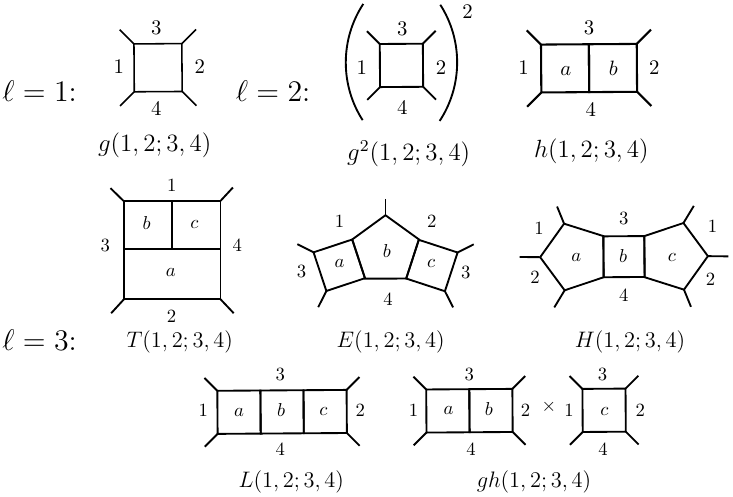} 
\centering
\caption{Topologies of correlator integrands up to $\ell=3$. The integrand $E^\prime(1,2;3,4), $ collapsing the propagator $x_{b,4}^2$ of $E(1,2;3,4)$, does not appear in the correlator but would be used in the chamber's loop-form.}
\label{fig:integrands}
\end{figure}

We now proceed with the loop-forms at one and two-loops. At one-loop, we introduce a pair of twistors $(Z_{A}, Z_{B})$ which can be expanded on any set of four twistors in $Z_{1,i}$ and $Z_{2,i}$. For example, we have 
\begin{equation}
\begin{pmatrix}
Z_A \\ Z_B
\end{pmatrix}=\begin{pmatrix}
1& x& 0 & -w \\ 0& y& 1& z
\end{pmatrix}\begin{pmatrix}
Z_{1,i} \\ Z_{2,i}\\ Z_{1,j}\\ Z_{2,j}
\end{pmatrix}
\end{equation}
Then for any given point in $\mathbb{T}_4$ we can construct a canonical form (4-form at one-loop) from the region defined in eq.(\ref{eq: LoopRegion}). Since the boundary of the region simply correspond to propagators, the loop-form can be naturally expanded on local integrands. The topologies relevant up to three loops are shown in FIG.~\ref{fig:integrands}.

At one loop, it turns out the loop-form for the two branches is nothing but the four-mass box integral normalized by (inverse) leading singularity $\pm \Delta$ for the branch $+$ and $-$ respectively
\begin{equation}
\Omega^{(1)\pm}=\frac{\pm\Delta~d^4 x_a}{x_{a,1}^2 x_{a,2}^2 x_{a,3}^2 x_{a,4}^2}\,, 
\end{equation}
where the loop variable is denoted as $x_a$. After combining with the form for the two branches, the complete one-loop form is then
\begin{equation}
\sum_{\pm}\;\omega^\pm\Omega^{(1)\pm} =\langle \! \langle \mathbf{X}_1, \mathbf{X}_2, \mathbf{X}_3, \mathbf{X}_4\rangle \! \rangle^4\frac{g(1,2,3,4)}{stu} \,,  
\end{equation}
where $g(1,2,3,4)$ represents the one-loop box integral. Thus we see that $\Delta$s, which contains square root, is canceled and we recover the well-known result of four-mass box integral. Note that the combination $\frac{g(1,2,3,4)}{stu}$ is permutation invariant with respect to all five points $x_i$ with $i=1,2,3,4,a$.

Moving to two loops, the loop-forms are expressed as linear combinations of the square of one-loop box and the double-box integral, which we denote as 
\begin{eqnarray}\label{eq: IntDef}
g(1,2,3,4)^2:=& \frac{d^4 x_a d^4 x_b}{2 x_{a,1}^2 x_{a,2}^2 x_{a,3}^2 x_{a,4}^2 x_{b,1}^2 x_{b,2}^2 x_{b,3}^2 x_{b,4}^2 
} + (a\leftrightarrow b)\,,\nonumber\\   
h(1,2;3,4):=& \frac{d^4 x_a d^4 x_b x_{3,4}^2}{x_{a,1}^2 x_{a,3}^2 x_{a,4}^2 x_{a,b}^2 x_{b,2}^2 x_{b,3}^2 x_{b,4}^2} + (a\leftrightarrow b)\,, 
\end{eqnarray}
where we have symmetrized the loop variables $a, b$; while $g(1,2,3,4)$ is permutation invariant with respect to the $4$ external points, $h(1,2; 3,4)$ is symmetric under the exchange of $1$ and $2$, as well as $3$ and $4$, so there are in total $6$ inequivalent permutations. What we find is that the two-loop forms for $+$ and $-$ regions are
\begin{equation}
\Omega^{(2)\pm}=\Delta^2 g(1,2,3,4)^2 \pm \Delta \left(h(1,2; 3,4) + 5~{\rm perm.}\right) 
\end{equation}
The sum of two branches again produces the correct integrand for two-loop correlators (with $\Delta$ canceled):
\begin{eqnarray}
&\sum_{\pm}\;\omega^\pm\Omega^{(2)\pm}=\frac{\langle \! \langle \mathbf{X}_1, \mathbf{X}_2, \mathbf{X}_3, \mathbf{X}_4\rangle\!\rangle^4}{stu}\times \nonumber\\
&\left((s{+}t{+}u)g(1,2,3,4)^2+ \left(h(1,2; 3,4) + 5~{\rm perm.}\right) \right)
\end{eqnarray}
once again the integrand is proportional to $\langle\! \langle \textbf{X}_1, \textbf{X}_2, \textbf{X}_3, \textbf{X}_4\rangle\!\rangle^4$ and the combination in the bracket is in fact permutation invariant with respect to six points $x_i$ with $i=1,2,3,4,a,b$.

\section{Three-loops and chambers}
\begin{figure}[H]
\includegraphics[width=8.5cm]{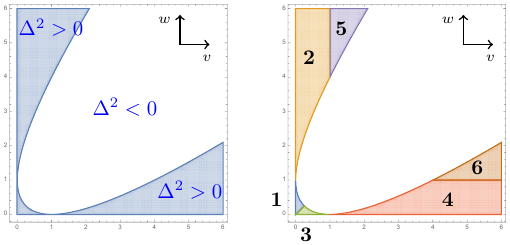} 
\centering
\caption{Shaded region (left) shows positive kinematics where $\Delta^2 > 0$. It splits into six chambers for $\ell=3$ (right). Axes are cross-ratios $v$ and $w$.}
\label{fig:3L-chambers}
\end{figure}

Starting from $\ell=3$, we find that the loop-form becomes dependent on kinematics. In particular, besides the two branches, $\mathbb{T}_4$ can be dissected into six different \textit{chambers} depending on the ordering of $s,t,u$, depicted in FIG.~\ref{fig:3L-chambers}:
\begin{equation}\label{eq:L=3_chambers}
    \begin{split}
        r_1:&\,s<t<u,\quad r_2:\,s<u<t,\quad r_3:\,t<s<u,\\
        r_4:&\,t<u<s,\quad r_5:\,u<s<t,\quad r_6:\,u<t<s\,.
    \end{split}
\end{equation}
The forms for these chambers can be computed straightforwardly. For example for $r_1$ we have
\begin{equation}\label{eq: ChamberForm}
 \omega_{r_1}^\pm=\frac{\langle \! \langle \mathbf{X}_1, \mathbf{X}_2, \mathbf{X}_3, \mathbf{X}_4\rangle\!\rangle^4}{\Delta^2}\times\left( \frac{1}{s (t-s)} {\pm} \frac{\Delta}{s(t-s)(u-t)}\right)\,.
\end{equation}
This form reflects the boundary of the chamber, $s=0$, $s=t$ and $t=u$. One obtains the form for other chambers by permuting $s,t,u$. Summing over the six chambers, we can recover the tree-form in eq.\eqref{eq:tree-form}.

\paragraph{Loop forms and three-loop integrands}
The loop-form for each chamber is now distinct, and we find a remarkably simple formula for all the six of them:
\begin{equation}\label{eq: 3Loops}
\Omega^{(3),\pm}_{r_i}= \Delta^2 A_{\sigma_3} \pm \Delta\big(B-\sigma_1 (C_{\sigma_2} + C_{\sigma_3}) - \sigma_2 C_{\sigma_1} \big)\,,   
\end{equation}
where the $\sigma_i$'s appear in the inequality $\sigma_1<\sigma_2<\sigma_3$ for each chamber in eq. (\ref{eq:L=3_chambers}). Thus $A_{\sigma_3}$ for each chamber is labeled by the largest $s,t,u$, and similarly for $C_{\sigma_i}$. The building blocks above contain the following integrals:
\begin{equation}\label{eq: block_function}
\begin{split}
         A_s:=&\big[H({1,4;2,3}){-} {E^{\prime}}({1,4;2,3})+ (1,4){\leftrightarrow}(2,3) \big]+ (3{\leftrightarrow} 4)\\
         & + gh({1,2;3,4})+ gh({3,4;1,2})\,,\\
    B:=& T({1,2;3,4}) +E({1,2;3,4}) + 11\, \text{perms.} \\
         & +L({1,2;3,4}) +(t+u) {E^{\prime}}({1,2;3,4}) + 5\, \text{perms.}\,,\\
C_s:=& 4 (E'(1, 2; 3, 4) + E'(3, 4; 1, 2))\\
\end{split}
\end{equation}
with the explicit integrals given in supplementary material B and the functions with subscript $t,u$ are defined as $\mathcal{O}_t=\mathcal{O}_{s}(2\leftrightarrow 3)$ and $\mathcal{O}_u=\mathcal{O}_{s}(2\leftrightarrow 4)$. For instance, the loop-form for the chamber $s<t<u$ is given by
\begin{equation}
    \Omega^{(3)\pm}_{r_1} =\Delta^2 A_{u} \pm \Delta \big(B- s\,(C_t+ C_u)- t\,C_s \big)\,.
\end{equation}
Very nicely, by summing over the $6$ chambers, of the chamber-form times loop-form, $\sum_{i,\pm}\;\omega_i^\pm \Omega_{r_i}^{(3)\pm}$, one reproduces exactly the three-loop answer in~\cite{EDEN2012450}.

\paragraph{Positive maximal cut}
The way the local integrals appear in the loop-form for each chamber can be understood via the positivity of maximal cuts. Since by construction, our loop-form can be expanded on a local integrand basis, if a given topology is present, that implies that its maximal cut must be positive in the sense that the cut (loop) solution satisfies eq.(\ref{eq: LoopRegion}). Such criteria was also utilized in the Landau singularity analysis for $\mathcal{N}=4$ SYM~\cite{Dennen:2016mdk}.  

All maximal cuts associated with topologies $T$, $E$, $L$ and $E^\prime$ are always positive, hence they are present in every chamber, i.e. the $B$ building blocks in eq.(\ref{eq: 3Loops}).
The maximal cut of typologies $H$, $gh$ are only positive in the specific region of positive kinematic. For example, the cut in $H(1,2;3,4)$, $gh(1,2;3,4)$ are positive only when $s$ is maximal, and thus appearing in $A_s$ building block. The presence of $E^\prime$ is more subtle as while its maximal cut is positive for all kinematics, the value of the cut changes sign as one crosses the boundaries of the chambers. This accounts for its appearance being dressed with appropriate factors of $s, t$, and $u$ in front of $C_{\sigma_i}$ building block.


\paragraph{Leading singularities from chamber's form}
As the loop-form for each chamber is built as a canonical form, this implies that the combination of integrands in~eq.\eqref{eq: 3Loops} are integrated to pure functions. Indeed in~\cite{Drummond:2013nda}, the three-loop result is given in combinations of logarithms and multiple polylogarithms, dressed with rational prefactors of $c_{1,2}$ that are leading singularities, i.e., the maximal residue of the associated integrals. Our result indicate that these leading singularities are in fact linear combinations of the chamber forms.

It is straightforward to verify this. The leading singularities that appear include 
\begin{gather}
    \frac{1}{ s t u \Delta},\ \frac{1}{(s{-}t)t u \Delta}, \ \frac{s+t}{s t u(s{-}t)t \Delta},\ \frac{1}{ s t  \Delta^2},\ \frac{s+t}{ s t u  \Delta^2}\quad 
\end{gather}
as well as their permutation with respect to $s,\ t,\ u$. Indeed they can all be written as linear combinations of eq.(\ref{eq: ChamberForm}). For example, we have
\begin{equation}
    \frac{s+t}{s t u(s-t)t \Delta}=\frac{1}{2}\sum_{\pm}\pm\left(\sum_{i=3,4,6}\omega_{r_i}^\pm-\sum_{i=1,2,5}\omega_{r_i}^\pm\right).
\end{equation}
Other leading singularities can be similarly represented.

\section{Loop forms on boundaries and lightlike limit}

 The loop-forms in each chamber must satisfy global consistency conditions. Given two chambers that share a boundary, their loop-form must agree on the boundary For example, the chambers $r_1, r_3$ share the boundary $s=t$, and indeed one finds:
\begin{equation}
\Omega^{(3)\pm}_{r_1}\Big|_{s=t}=\Omega^{(3)\pm}_{r_3}\Big|_{s=t}\,.
\end{equation}
On the other hand, if the boundary shared by two chambers resides in the region $\Delta^2<0$, then the loop-form should vanish. For example, chamber $1$ and $2$ share the boundary  $s<t=u$, which leads to  $\Delta^2\big|_{t=u}=s(s-4u)<0$. Indeed on the boundary:
\begin{equation}
    \Omega^{(3)\pm}_{r_1}\Big|_{t=u}=\Omega^{(3)\pm}_{r_2}\Big|_{t=u}=0\,.
\end{equation}

The matching of loop-forms on the boundary also has consequences on the lightlike limit of the geometry, where the consecutive points become null separated. It is known that the correlator then becomes dual to the square of the amplitude~\cite{Eden:2010ce, Eden:2011ku, Alday:2010zy, Eden:2010zz, Eden:2011yp, Adamo:2011dq}. We can take the lightlike limit by setting the bitwistors to $X^{IJ}_i=Z^{[I}_i Z^{J]}_{i{+}1}$ with $Z_5=Z_1$. In this limit $s, t\rightarrow 0$ and $u\rightarrow 1$ thus $\mathbb{T}_4$ collapse to a point. The geometry then becomes the same as the four-point Amplituhedron scans sign flipping conditions. As discussed in~\cite{Dian:2021idl}, this geometry indeed corresponds to the square of Amplituhedron.

\section{Conclusions and outlook}
We have considered a positive geometry that captures the all-loop planar correlator of $\mathcal{N}=4$ SYM. The construction utilizes the notion of fibration, where one defines a positive space on which a loop-geometry is attached to each point, and the latter in turn dissects the base in distinct chambers. We have verified the resulting loop-form combined with chamber-form matches to known results up to three loops. An immediate question is whether the chambers that emerged at three loops suffice for dissecting the all-loop geometry. In a separate work~\cite{longpaper} we will provide evidence this is indeed the case at four loops. This might be somewhat puzzling, as some four-loop integrals are known to contain elliptic polylogarithms (whose elliptic curves can be seen from the corresponding maximal cuts). Yet our construction implies that at any loop order, the answer should be given by a product of chamber forms which are at most algebraic, with some pure functions from integrating $d\log$ forms. It would be interesting to clarify this, and to understand more generally the meaning of chambers for both the correlator and higher-point amplitudes in SYM/ABJM. 

It is straightforward to generalize our construction to higher points, which at least reduces to the correct squared amplituhedron~\cite{Dian:2021idl}. It would be highly desirable to see if it continues to give the correct loop geometry for correlators and the nature of higher-point chambers. Obviously, one can make contact with the original proposal of correlahedron~\cite{Eden:2017fow} and understand the precise relation. The separation of tree and loop-geometry also makes it natural to decompose the latter into negative geometries  originally introduced in~\cite{Arkani-Hamed:2021iya} (further discussed in~\cite{Chicherin:2022zxo,Chicherin:2022bov,Brown:2023mqi,He:2023exb,Henn:2023pkc,Lagares:2024epo,Li:2024lbw}), making it highly desirable to study loop integrations for these negative geometries of the correlators and even consider possible all-loop resummations. Moreover, given the hidden ten-dimensional symmetry presented in all-loop integrands~\cite{Caron-Huot:2021usw}, it would be extremely interesting to consider extending our geometries to incorporate correlators involving the infinite tower of half-BPS operators. Finally, by construction our geometry is permutation invariant with respect to the four external points, reflecting the properties of the correlation function. It will be interesting to break it down to cyclic invariant, while keeping the external points spacelike separated. This would be the desired property of Coloumb branch amplitudes, and might be closely related to the construction of ~\cite{Arkani-Hamed:2023epq}.

\begin{acknowledgments}
We thank  Gabriele Dian, Paul Heslop, and Congkao Wen for very informative discussions on this project. The work of S.H. is supported by the National Natural Science Foundation of China under Grant No. 12225510, 11935013, 12047503, 12247103, and by the New Cornerstone Science Foundation through the XPLORER PRIZE. The research of Y.-t. Huang and C.-K. K. are supported by the Taiwan Ministry of Science and Technology Grant No. 112-2628-M-002 -003 -MY3. Additionally, C.-K. K. is also funded by the ERC Grant No. 101118787, UNIVERSE PLUS, with the views expressed being those of the authors and not necessarily reflecting those of the EU or ERC, which are not responsible for them.
\end{acknowledgments}

\bibliographystyle{apsrev4-1}
\bibliography{bib}

\newpage

\appendix

\widetext
\begin{center}
\textbf{\Large Supplementary Material}
\end{center}
\section{Conventions and the universal factor for four-point correlator}\label{sec: universal}
We consider four-point correlation function in planar ${\cal N}=4$ SYM with the simplest half-BPS operator ${\cal O}_{20'}^{IJ}:=tr(\Phi^I \Phi^J)-\frac 1 6 \delta^{IJ} tr(\Phi^K \Phi^K)$, and introduce the auxiliary ``harmonics" $y$-variables (for more details refer to~\cite{Eden:2011we, Chicherin:2014uca})
\begin{equation}
G_4=\langle {\cal O}(x_1 y_1) \cdots {\cal O}(x_4, y_4)\rangle=\sum_{L=0}^\infty a^{\ell} G_4^{(\ell)},
\end{equation}
where $a=g^2 N_c/(4\pi^2)$ is the t' Hooft coupling, and the tree correlator $G_4^{(0)}$ is given by 
\begin{equation}
G_4^{(0)}=\frac{(N_c^2-1)^2}{(4\pi^2)^4}(d_{12}^2 d_{34}^2 + 2~{\rm perms.}) + \frac{N_c^2-1}{(4\pi^2)^2}(d_{12} d_{23} d_{34} d_{41} + 2~{\rm perms.})
\end{equation}
where $d_{i,j}:=\frac{y_{i,j}^2}{x_{i,j}^2}$ with $y_{i,j}=(y_i-y_j)^2$ from harmonics. Starting $\ell>1$, the correlator is given by a universal prefactor, $R(1,2,3,4)$, times the $\ell$-loop function $F^{(\ell)}$:
\begin{equation}
G_4^{(\ell)}=\frac{2 (N_c^2-1)}{(4\pi^2)^4}\,R(1,2,3,4)\times F^{(L)}(x_1, x_2, x_3, x_4)\qquad {\rm for}~\ell\geq 1 
\end{equation}
and we are interested in its {\it integrand} which also depends on the loop variables $x_a\equiv x_5, x_b\equiv x_6, \cdots, x_{4+\ell}$:
\begin{equation}
F^{(\ell)}:=\frac{\prod_{1\leq i<j\leq 4} x_{i,j}^2}{\ell!} \int \prod_{a=1}^\ell d^4 x_{4{+}a} I^{(\ell)}(x_1, \cdots, x_{4{+}\ell})\,.
\end{equation}
As explained in details in~\cite{Eden:2011yp}, this integrand can be extracted from the correlation function with $\ell$ Lagrangian insertions and enjoys a remarkable permutation invariance among all $4+\ell$ points. We conjecture that the all-loop geometry encodes the complete information about $I^{(\ell)}$, and let us first explain the appearance of the universal prefactor. 

In~\cite{Chicherin:2014uca} the supersymmetric extension was constructed in chiral superspace $(x^{\alpha \dot{\alpha}},\theta^{I\alpha})$, where $\mathcal{O}(x_i, \theta_i)$ is the super-BPS operator and the correlator is extracted from a potential $\mathcal{G}(x,\theta)$ via 
\begin{equation}
\langle \mathcal{O}(x_1,\theta_1)\mathcal{O}(x_2,\theta_2)\mathcal{O}(x_3,\theta_3)\mathcal{O}(x_4,\theta_4)\rangle=\left[\prod_{i=1}^4 D^4_i\right]\mathcal{G} (x,\theta)
\end{equation}
with differential operators $D^4_i$ defined as,
\begin{equation}\label{eq: D4Def}
D^4_i=y_{i}^{IJ} y_{i}^{KL} \frac{\partial}{\partial \theta_i^{I\alpha}}\frac{\partial}{\partial \theta_i^{J\beta}}\frac{\partial}{\partial \theta_{i\alpha}^K}\frac{\partial}{\partial \theta_{i\beta}^L}\,,
\end{equation}
where in our convention the $4\times 4$ matrices $y_i$ are given by
\begin{equation}
    y_i^{IJ}=
\begin{pmatrix}
\epsilon_{2\times 2} & (y_i)_{2\times 2} \\
-(y_i)_{2\times 2}^T &  \det y_i\,\epsilon_{2\times 2}
\end{pmatrix}\,,\quad
    (y)_{2\times 2}=
\begin{pmatrix}
y^1-y^4 & y^2-y^3 \\
y^2+y^3 & y^1+y^4
\end{pmatrix}\,.
\end{equation}

Our geometry gives the potential $\mathcal{G} (x,\theta)$ where the fermionic variables are bosonized~\cite{Arkani-Hamed:2013jha}. More precisely, We introduce the bosonied bitwistors:
\begin{equation}
    \textbf{X}_i=
\begin{pmatrix}
Z_{i,\alpha} \\
\theta_i^\alpha \cdot \phi_1\\
\vdots\\
\theta_i^\alpha \cdot \phi_4
\end{pmatrix}\,.
\end{equation}
Our potential is then a function of $Y$ and $\mathbf{X}$, $\mathcal{G}(\mathbf{X},Y)$. The potential in chiral superspace is given by
\begin{equation}
\mathcal{G}(x,\theta)=\int d^{16}\phi\;\mathcal{G}(\mathbf{X},Y)\bigg|_{\llangle Y, \mathbf{X}_i, \mathbf{X}_j\rrangle\rightarrow x^2_{i,j}}
\end{equation}
Note that after identifying $\llangle Y, \mathbf{X}_i, \mathbf{X}_j\rrangle\rightarrow x^2_{i,j}$ our potential will only depend on bosonized twistor through the universal factor $\langle \! \langle \mathbf{X}_1, \mathbf{X}_2, \mathbf{X}_3, \mathbf{X}_4\rangle \! \rangle^4$. We have
\begin{equation}
    \int d^{16}\phi\, \langle \! \langle \mathbf{X}_1, \mathbf{X}_2, \mathbf{X}_3, \mathbf{X}_4\rangle \! \rangle^4=X^\perp \cdot \Theta^\mathcal{I}
\end{equation}
where the column vector $\Theta^\mathcal{I}=\left\{ \theta_1^{\mathcal{I}1}, \theta_1^{\mathcal{I}2}, \cdots, \theta_4^{\mathcal{I}1}, \theta_4^{\mathcal{I}2} \right\}$. Acting on the result with the fermionic derivatives as in \eqref{eq: D4Def} one finally arrives at
\begin{equation}
    \left(\prod_{i=1}^4 D_i\right) X^\perp \cdot \Theta^\mathcal{I}=R(1,2,3,4)\prod_{i<j} x_{i,j}^2
\end{equation}
where we have obtained the desired universal prefactor as
\begin{eqnarray}
R(1,2,3,4):= \left(d_{1,2} d_{2,3} d_{3,4} d_{4,1} (1{-}u{-}v){+}d_{1,3}^2 d_{2,4}^2\right) x_{1,3}^2 x_{2,4}^2  
+(1\leftrightarrow 2) + (1\leftrightarrow 4) \nonumber \\    
\end{eqnarray}

\section{Conformal integrals for three loops}\label{appendix: Int}
Here we collect the definition of conformal integrals, which together with their permutations wrt $1,2,3,4$ form a integral basis for computing the three-loop forms. Note that $T, E, H, L$ and $gh$ are those appearing in the final answer (see FIG.~\ref{fig:integrands}), and we also need $E'(1,2;3,4)$ (and its permutations) for the loop form at $\ell=3$, which is obtained by collapsing the propagator $x_{b,4}^2$ of $E(1,2;3,4)$. 
\begin{equation}
    \begin{split}
        T(1,2;3,4)&:=\frac{d^4 x_a d^4 x_b d^4 x_c x_{a,1}^2 x_{3,4}^2}{x_{a,2}^2x_{a,3}^2 x_{a,4}^2 x_{a,b}^2 x_{b,1}^2 x_{b,3}^2 x_{b,c}^2 x_{c,1}^2 x_{c,4}^2 x_{a,c}^2 } + \text{perms}(a,b,c)\\
        E(1,2;3,4)&:=\frac{d^4 x_a d^4 x_b d^4 x_c x_{b,3}^2 x_{1,4}^2 x_{2,4}^2}{x_{a,1}^2 x_{a,3}^2 x_{a,4}^2 x_{a,b}^2 x_{b,1}^2 x_{b,2}^2 x_{b,4}^2 x_{b,c}^2 x_{c,2}^2 x_{c,3}^2 x_{c,4}^2}+ \text{perms}(a,b,c)\\
        E^\prime(1,2;3,4)&:=\frac{d^4 x_a d^4 x_b d^4 x_c}{x_{a,1}^2 x_{a,3}^2 x_{a,4}^2 x_{a,b}^2 x_{b,1}^2 x_{b,2}^2  x_{b,c}^2 x_{c,2}^2 x_{c,3}^2 x_{c,4}^2}+ \text{perms}(a,b,c)\\
        L(1,2;3,4)&:=\frac{d^4 x_a d^4 x_b d^4 x_c x_{3,4}^4}{x_{a,1}^2 x_{a,3}^2 x_{a,4}^2 x_{a,b}^2 x_{b,3}^2 x_{b,4}^2  x_{b,c}^2 x_{c,2}^2 x_{c,3}^2 x_{c,4}^2}+ \text{perms}(a,b,c)\\
        H(1,2;3,4)&:=\frac{d^4 x_a d^4 x_b d^4 x_c x_{a,c}^2 x_{1,3}^2}{2\,x_{a,1}^2 x_{a,2}^2 x_{a,3}^2 x_{a,4}^2 x_{a,b}^2 x_{b,3}^2 x_{b,4}^2  x_{b,c}^2 x_{c,1}^2x_{c,2}^2 x_{c,3}^2 x_{c,4}^2}+ \text{perms}(a,b,c)\\
        gh(1,2;3,4)&:=\frac{d^4 x_a d^4 x_b   x_{3,4}^2}{x_{a,1}^2  x_{a,3}^2 x_{a,4}^2 x_{a,b}^2 x_{b,2}^2 x_{b,3}^2 x_{b,4}^2   }\times \frac{d^4 x_c}{x_{c,1}^2x_{c,2}^2 x_{c,3}^2 x_{c,4}^2}+ \text{perms}(a,b,c)
    \end{split}
\end{equation}

\end{document}